\documentclass[9pt,conference]{IEEEtran}
\IEEEoverridecommandlockouts

\usepackage{cite}
\usepackage{float}
\usepackage{amsmath,amssymb,amsfonts}
\usepackage{algorithmic}
\usepackage{graphicx}
\usepackage{textcomp}
\usepackage{xcolor}
\usepackage[table]{xcolor}
\def\BibTeX{{\rm B\kern-.05em{\sc i\kern-.025em b}\kern-.08em
    T\kern-.1667em\lower.7ex\hbox{E}\kern-.125emX}}

\begin{document}

\title{In-Pipeline Integration of Digital In-Memory-Computing into RISC-V Vector Architecture to Accelerate Deep Learning}














\author{%
\IEEEauthorblockN{%
Tommaso Spagnolo\IEEEauthorrefmark{1} \quad
Cristina Silvano\IEEEauthorrefmark{1}\\[0.5ex]
Riccardo Massa\IEEEauthorrefmark{2} \quad
Filippo Grillotti\IEEEauthorrefmark{2} \quad
Thomas Boesch\IEEEauthorrefmark{3} \quad
Giuseppe Desoli\IEEEauthorrefmark{2}}
\IEEEauthorblockA{\IEEEauthorrefmark{1}Politecnico di Milano, Milan, Italy\\
Email: tommaso.spagnolo@polimi.it, cristina.silvano@polimi.it}
\IEEEauthorblockA{\IEEEauthorrefmark{2}STMicroelectronics, Milan, Italy\\
Email: riccardo.massa@st.com, filippo.grillotti@st.com, giuseppe.desoli@st.com}
\IEEEauthorblockA{\IEEEauthorrefmark{3}STMicroelectronics, Geneva, Switzerland\\
Email: thomas.boesch@st.com}
}


\maketitle

\IEEEaftertitletext{\vspace{10\baselineskip}}

\begin{abstract}
Expanding Deep Learning applications toward edge computing demands architectures capable of delivering high computational performance and efficiency while adhering to tight power and memory constraints. Digital In-Memory Computing (DIMC) addresses this need by moving part of the computation directly within memory arrays, significantly reducing data movement and improving energy efficiency. 
This paper introduces a novel architecture that extends the Vector RISC-V Instruction Set Architecture (ISA) to integrate a tightly coupled DIMC unit directly into the execution stage of the pipeline, to accelerate Deep Learning inference at the edge. Specifically, the proposed approach adds four custom instructions dedicated to data loading, computation, and write-back, enabling flexible and optimal control of the inference execution on the target architecture.
Experimental results demonstrate high utilization of the DIMC tile in Vector RISC-V and sustained throughput across the ResNet-50 model, achieving a peak performance of 137 GOP/s. 
The proposed architecture achieves a speedup of 217× over the baseline core and 50× area-normalized speedup even when operating near the hardware resource limits. 
The experimental results confirm the high potential of the proposed architecture as a scalable and efficient solution to accelerate Deep Learning inference on the edge.
\end{abstract}

\begin{IEEEkeywords}
AI Accelerator, RISC-V Architecture, Instruction Set Extension, Digital In-Memory Computing, Vector Processors
\end{IEEEkeywords}

\section{Introduction}
Artificial Intelligence continues to reshape technology and society, driven by the rapid evolution of increasingly complex AI models. Since the advent of modern Deep Learning with AlexNet in 2012 \cite{krizhevsky_imagenet_2012}, AI model complexity, typically measured by parameter count, has roughly doubled each year \cite{noauthor_data_2024-1}. Conversely, hardware performance has improved at a substantially slower pace, achieving only around 30\% annual growth in terms of FP32 TFLOP/s \cite{noauthor_data_2024}. This widening gap has made it clear that traditional computing architectures can no longer rely solely on incremental improvements; instead, architectural innovation is required to effectively manage the computational demands of emerging AI workloads. Furthermore, contemporary AI models have become increasingly diverse, encompassing various neural network types, evolving inter-layer connectivity, distinct data-flow patterns, diverse activation functions, and multiple numerical formats (e.g., quantization levels). These variations place significant emphasis on computational flexibility, as fixed-function hardware accelerators, though efficient, often lack the adaptability required to accommodate such rapidly changing and heterogeneous workloads.

The widespread adoption of AI has highlighted significant drawbacks in centralized cloud computing, especially regarding latency, energy efficiency, and data security \cite{silvano_survey_2024}. These challenges have driven a shift toward edge computing, where AI inference occurs locally on devices, providing lower latency, improved security, and reduced dependency on network connectivity. However, edge deployment imposes strict constraints on power, memory, and performance, necessitating specialized hardware capable of high computational throughput within tight energy and memory limits.

At the device level, the primary obstacle to efficiently executing advanced AI workloads arises from excessive data movement rather than computation itself. Modern Deep Learning architectures, such as Convolutional Neural Networks (CNNs) and transformers, frequently move large volumes of feature maps, weights, and intermediate results across memory hierarchies, placing immense pressure on memory bandwidth. Traditional digital accelerators often struggle with significant latency and energy overheads, largely driven by continuous data transfer between memory and compute units. Indeed, transferring data from off-chip DRAM typically consumes orders of magnitude more energy than performing actual Multiply-Accumulate (MAC) operations \cite{horowitz_11_2014}, primarily due to interconnect capacitance.

\begin{figure}[!b]
    \centering
    \includegraphics[width=\columnwidth]{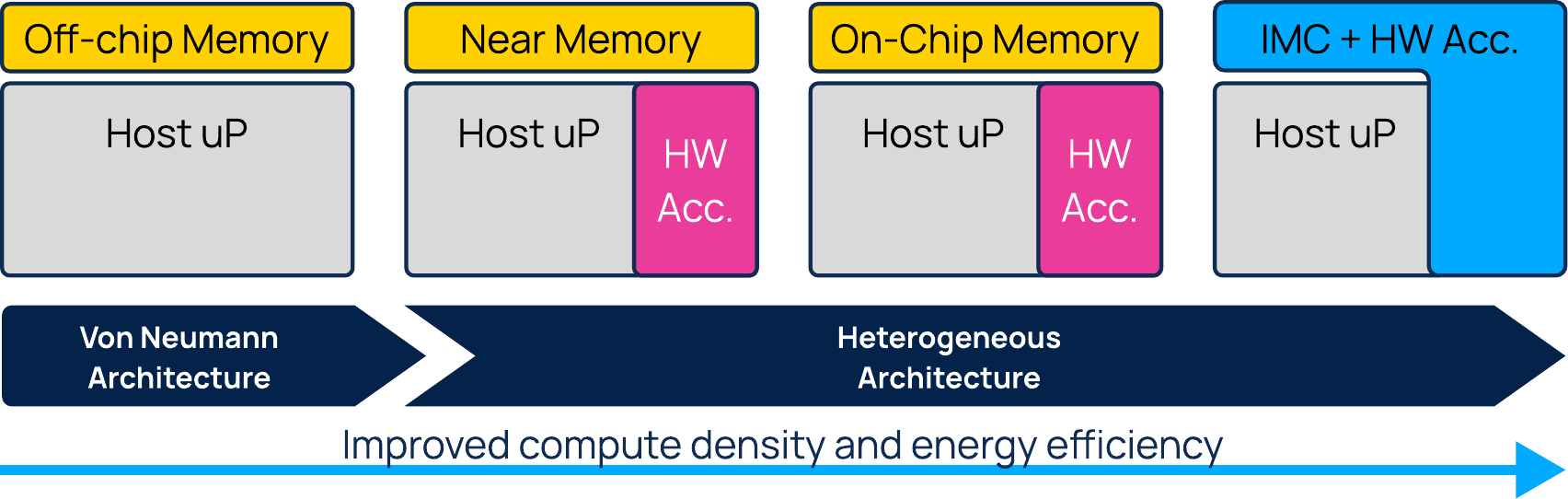}
    \caption{Architectural approaches to improve data locality: from von Neumann designs to heterogeneous systems with near-memory and in-memory compputing.}
    \label{fig:data_locality}
\end{figure}

Addressing this critical bottleneck requires improving data locality and reducing the physical distance between memory and computation (see Fig.~\ref{fig:data_locality}). Near-memory computing has emerged as a promising strategy, which involves placing small, high-speed memories, such as scratchpads, close to compute units. 
By storing frequently accessed data near the compute elements, these architectures significantly minimize energy consumption and latency, promoting data reuse.

Building on the concept of near-memory computing, In-Memory Computing (IMC) takes this paradigm even further by integrating computational capabilities directly in memory.


IMC architectures are classified into Analog (AIMC) and Digital (DIMC). AIMC uses emerging Non-Volatile Memory (NVM) technologies like ReRAM or PCM to perform analog MACs \cite{kim_overview_2024}, achieving energy efficiency over 1000 TOPS/W at low precision \cite{hung_8-mb_2022}, but suffers from reliability and accuracy issues \cite{ielmini_-memory_2018}, limiting adoption in precision-critical tasks. DIMC, on the other hand, uses digital SRAM arrays and offers deterministic behavior, high bit-precision, and seamless CMOS integration. Though typically achieving lower energy efficiency (10–300 TOPS/W) than AIMC, DIMC ensures robustness and accuracy, making it ideal for fully-digital CNN accelerators at the edge \cite{desoli_167_2023, fujiwara_5-nm_2022}. 


Integrating IMC units into processors involves choosing between tightly-coupled and loosely-coupled configurations. In loosely-coupled designs, IMC tiles act as accelerators accessed via load/store instructions through the processor’s I/O bus at specific memory-mapped addresses. Although scalable and simple to integrate, this approach introduces significant communication overhead, causing stalls and performance loss. Conversely, tightly-coupled designs embed IMC units directly into the pipeline or nearby memory, leveraging ISA extensions for direct access. This reduces communication latency to single-digit cycles by avoiding memory hierarchy and interconnect delays, though it adds complexity through extra pipeline ports, hazard logic, tighter timing, and extensive verification. Despite this complexity, tightly-coupled integration provides superior responsiveness, efficiency, and flexibility for adapting to various neural network topologies and dynamic data flows.


The RISC-V ISA, with its open and modular structure, provides an ideal foundation for exploring these integration strategies. Its flexibility and modularity allow designers to introduce custom instructions that support IMC operation, whether tightly integrated or offloaded, while maintaining compatibility with existing toolchains.


Building on this modularity, the RISC-V vector extension \cite{noauthor_riscv-v-specv-specadoc_nodate} introduces native support for data-parallel workloads such as CNN inference, Digital Signal Processor (DSP), and computer vision algorithms. By exploiting Data-Level Parallelism (DLP), vector processors can achieve high throughput while maintaining energy and area efficiency—key advantages for edge AI. Unlike fixed-width Single Instruction-Multiple Data (SIMD) models, the vector extension supports variable-length registers, allowing developers to tailor vector operations to different applications and hardware budgets.

This flexibility and efficiency position RISC-V vector units as a strong alternative to traditional GPUs and multicore processors. Compared to the latter, vector units offer improved power efficiency and simpler programming models, while against GPUs, 
they deliver better area utilization and lower energy consumption. These features make vector architectures—especially in the RISC-V ecosystem—a compelling choice for building general-purpose yet efficient accelerators that integrate emerging paradigms like IMC.

Within this context, this work makes three main contributions:

\textbf{\textit{First,}} we present an efficient method for tightly coupling a DIMC unit within the vector processor pipeline. This integration enables the DIMC to be fully exploited by extending workload mapping flexibility at the assembly instruction level and an efficient utilization across a wide range of current and emerging workloads.

\textbf{\textit{Second,}} we propose a custom ISA extension to the RISC-V vector standard, designed to manage the DIMC integration efficiently. Four new vector instructions are proposed: two for loading data into the DIMC and two for managing the computation start and write-back operations. These instructions maximize the unit’s bandwidth and throughput while maintaining compatibility with the RISC-V vector instruction encoding, ensuring ease of reuse in future architectures.

\textbf{\textit{Third,}} we prove the effectiveness of our integration on an industrial-level RISC-V vector core. Experimental results on ResNet50 show that embedding a single DIMC tile yields up to 137 GOPS and over 200× speedup compared to the baseline, even under tight hardware constraints. This baseline-oriented comparison highlights the substantial incremental benefit enabled by our proposed integration method.

\textit{Paper Structure.} Section II reviews state-of-the-art RISC-V architectures integrating IMC. Section III details the proposed architectural design. Section IV introduces the ISA extension. Section V presents the experimental setup and performance results. Section VI concludes with a summary and future directions.

\section{State of the Art}

Recent research has explored the integration of In-Memory Computing (IMC) into RISC-V-based processors to improve the efficiency of Deep Learning inference. While both analog and digital IMC architectures have demonstrated significant potential, their integration within programmable processors, especially vector-capable ones, remains limited. This section reviews representative works that integrate IMC units into RISC-V general-purpose or vector processors, distinguishing between tightly and loosely coupled models. The analysis emphasizes reported performance in terms of peak throughput (GOPS or TOPS) and supported precision, and positions them with respect to our approach.

\subsection{Tightly Coupled IMC in RISC-V Cores}
Tightly coupled architectures integrate IMC units directly into the RISC-V pipeline, enabling low-latency communication and instruction-level control. For instance, \textbf{AI-PiM} \cite{verma_ai-pimextending_2022} extends the RV64IMC ISA with custom PiM instructions, integrating compute-in-memory units within the processor pipeline. This design achieves speedups of up to 17.63× in matrix-vector multiplication and an average improvement of 2.74× across MLPerf Tiny benchmarks. However, AI-PiM targets scalar pipelines, and therefore lacks the scalability required for highly parallel workloads.

A different approach is taken by \textbf{Vecim} \cite{wang_306_2024}, which integrates an 8T SRAM-based CIM architecture as a Vector Register File within a RISC-V vector core. Supporting multiple precision formats including INT8, BF16, and FP16, the architecture achieves a peak performance of 31.8 GOPS at 250 MHz. Nevertheless, while Vecim benefits from vector capabilities, its design primarily emphasizes register-file integration rather than functional unit extension, limiting its flexibility in supporting diverse computational patterns.

\textbf{RDCIM} \cite{10403109} instead presents a fully digital CIM processor tightly coupled with a RISC-V core through extended instructions for fine-grained control. It incorporates techniques such as Adding-on-Memory-Boundary (AOMB) and a Multi-Precision Adaptive Accumulator (MPAA) to reduce power and area overheads, supporting 4/8/12/16-bit precision with high energy efficiency. The design demonstrates 66.3 TOPS/W in 4-bit mode and 16.6 TOPS/W in 8-bit mode. Although RDCIM introduces valuable reconfigurable features, it remains focused on scalar pipelines, limiting its applicability to workloads requiring wide vector-level parallelism.

\subsection{Loosely Coupled IMC in RISC-V Cores}

Loosely coupled architectures extend RISC-V cores with IMC engines functioning as co-processors, communicating through buses or custom synchronization schemes. \textbf{VPU-CIM} \cite{j_vpu-cim_2024}, for example, integrates RRAM-based CIM with vector ISA extensions, achieving 33.98 TOPS/W with support for variable-bit precision (1–4 bits). However, since its CIM operations are decoupled from the main pipeline, this approach suffers from additional communication overhead and reduced fine-grained control.

Similarly, \textbf{CIMR-V} \cite{guo_cimr-v_2024} embeds a 10T SRAM-based digital CIM engine alongside a convolution and pooling pipeline, controlled via dedicated instructions. Fabricated in TSMC 28nm, it reaches 26.2 TOPS at 50 MHz for keyword spotting tasks, demonstrating the potential of IMC for domain-specific acceleration. Yet, the offloading model introduces latency and constrains flexibility in AI workloads.

\vspace{5pt}
In summary, prior works demonstrate either feasibility (scalar tight coupling) or energy efficiency (vector loose coupling), but none integrate a DIMC directly into the vector pipeline. Our approach fills this gap by embedding DIMC as a functional unit, enabling tight control, scalable vector processing, and efficient execution of complex dataflows while preserving programmability through the standard vector ISA.

\section{Proposed DIMC integration within the RISC-V Vector Pipeline}
\begin{figure}[!b]
    \centering
    \includegraphics[width=\columnwidth]{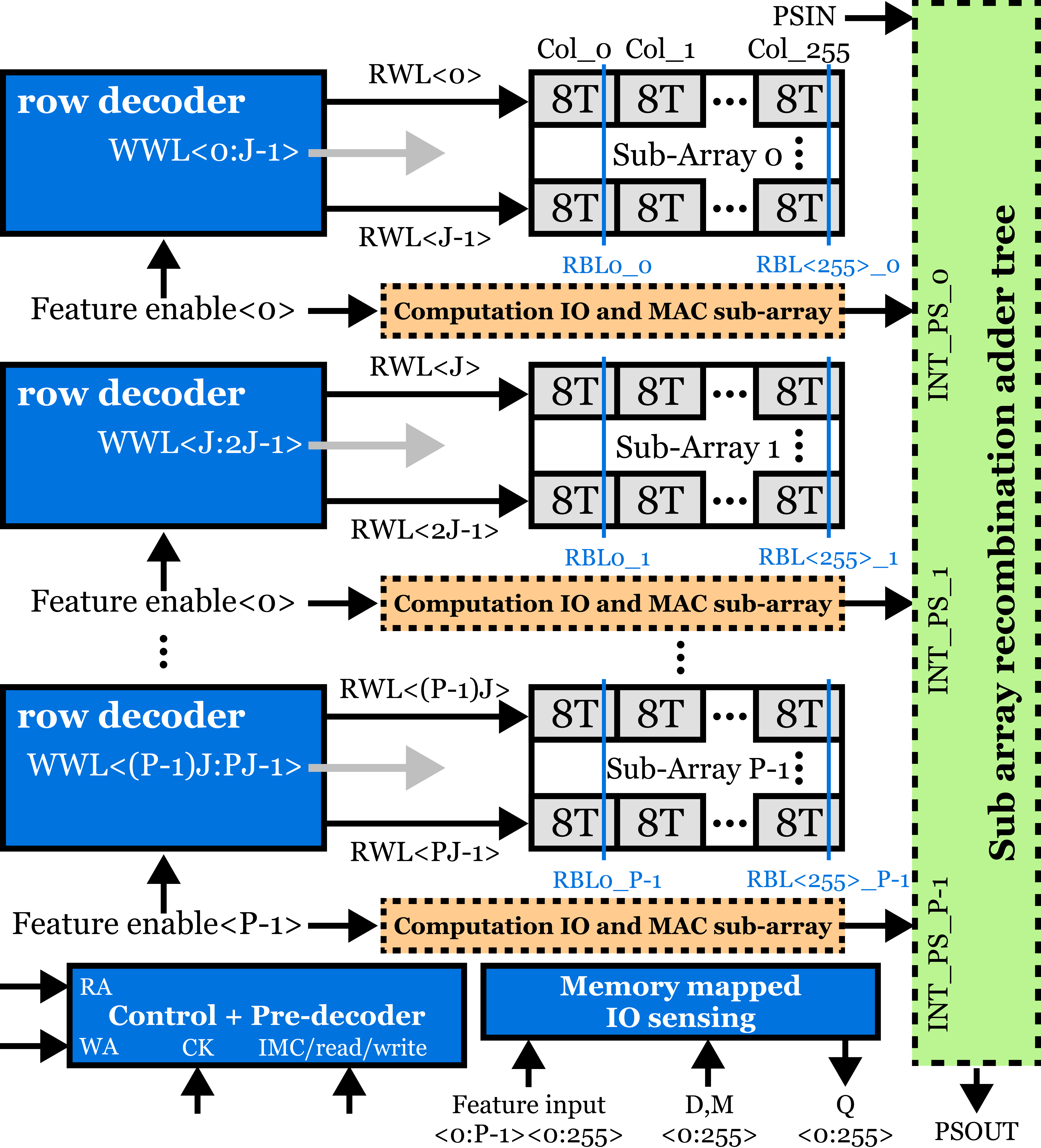}
    \caption{Architecture of DIMC Tile (P sub-arrays, J rows each) as presented at ISSCC 2023 \cite{desoli_167_2023}}
    \label{fig:desoli}
\end{figure}

The DIMC unit used in this work is the one presented in \cite{desoli_167_2023} and shown in Fig.~\ref{fig:desoli}. It features a memory capacity of 32 KiB, organized into 32 rows of 1024 bits each, typically used to store kernel weights during convolution execution. In addition, a 1024-bit input buffer is available to store the feature data used in the computation. The latter, takes place directly between the input buffer and a selected row of the memory, enabling in-memory execution of MAC operations and minimizing data movement overhead.

Internally, the DIMC architecture is structured into multiple sub-arrays, each consisting of 8T 1R1W bitcells and equipped with independent read word lines (RWLs) and read bitlines (RBLs). These sub-arrays can operate either as a unified array in memory-mapped mode or as independent units in compute mode. During computation, multiple RBLs are accessed in parallel within each sub-array, and the resulting signals are sensed and routed through dedicated IO paths to interleaved MAC slices.

This setup supports 256 parallel signed or unsigned 4-bit MAC operations per cycle, implemented as four parallel sub-arrays of 64 MACs each, all sharing a common accumulation pipeline. The architecture also supports runtime reconfiguration, allowing the same hardware to perform 512 2-bit or 1024 1-bit MAC operations per cycle, offering flexibility in precision. This configurability enables a scalable trade-off between accuracy and efficiency, making the DIMC ideal for edge AI inference.   
The memory interface supports 256-bit data transfers per cycle for both read and write operations. The computation produces 24-bit partial results, which can optionally be passed through a ReLU activation stage before being written back.

\begin{figure}[!b]
    \centering
    \includegraphics[width=\columnwidth]{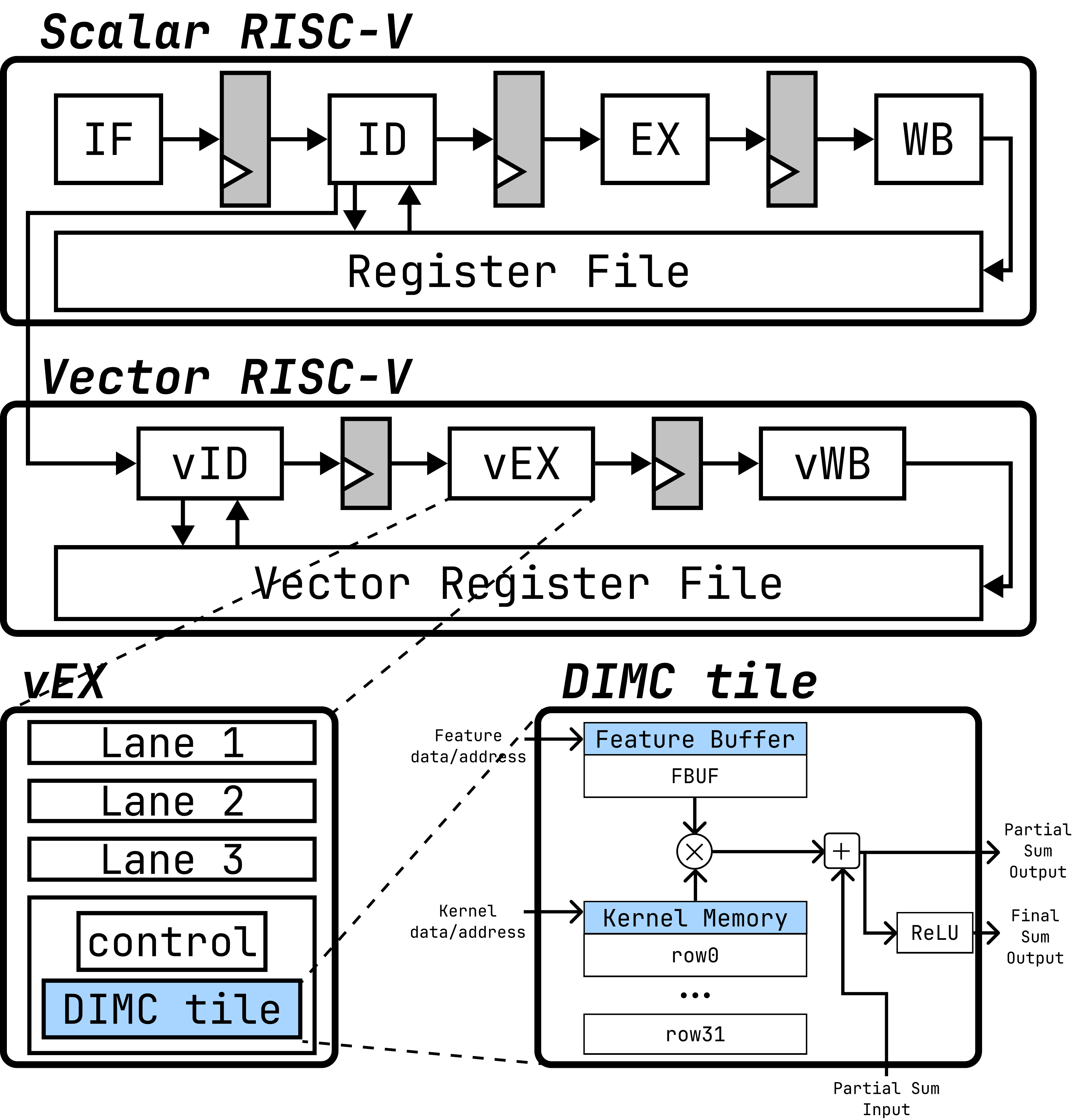}
    \caption{Architecture overview with the DIMC tile Integrated in the Vector Execution Stage as a Parallel Execution Lane}
    \label{fig:pipeline_DIMC_BIG}
\end{figure}

We integrate this DIMC as a dedicated execution lane in a RISC-V core implementing the embedded vector extension profile Zve32x, with architectural parameters VLEN = 64 and ELEN = 32, based on the industrial RVV implementation described in~\cite{bordonaro_integration_2025} (Fig.~\ref{fig:pipeline_DIMC_BIG}). The unit receives operands and decoded instructions like any other FU, but is driven by new custom vector instructions introduced in Sec.~IV.These instructions orchestrate high-bandwidth data loading, fine-grained configuration, and low-overhead result storage, all while allowing the DIMC lane to run in parallel with standard vector FUs.

\textbf{Why a vector core matters.} Beyond raw MAC throughput, overall performance hinges on how quickly feature maps and kernels can be folded, packed, or transposed into the 256-bit stripes the DIMC consumes. The RISC-V vector ISA already provides the data-manipulation primitives needed, and the VRF exposes sufficient read/write ports to match the DIMC bandwidth. These allow software to reshape irregular, dilated, grouped, or highly quantized tensors entirely within the VRF. Consequently, the vector core acts as a flexible “data-manipulator” that feeds a high-performance DIMC compute engine without extra glue hardware; the only cost is a modest cycle overhead inside the VRF. This synergy vastly broadens the range of convolution workloads the architecture can support.

This work concentrates on a single DIMC tile, with the primary goal of defining efficient integration and control within a vector core. Scaling to multiple tiles naturally follows as future work.

\section{Proposed RISC-V Instruction Set Extension}
Integrating a non-standard functional unit like the DIMC into a RISC-V vector core requires dedicated custom instructions. While a simple hardware connection may enable basic operation, fully leveraging the DIMC’s capabilities requires tight integration and instruction-level customization. This includes fine-grained control of data movement, execution order, and result handling to maximize in-memory convolution efficiency.

Although physically integrated into the Execution Stage of the pipeline, the DIMC unit operates under a specialized execution model that differs from standard vector FUs:
\begin{enumerate}
    \item Data loading: The DIMC must first be loaded with both weights and input data before computation can begin. 
    \item Computation: The DIMC performs in-memory MAC operations directly between the input buffer and selected memory rows, and requires explicit control over the target rows, applied masks, and computation precision.
    \item Result Write-Back: Partial sums or final outputs of each row are generated sequentially, one per cycle. If not properly synchronized, these results can be lost, making precise timing and control essential.
\end{enumerate}

To support this model, the custom instructions must be carefully designed. To this end, a custom RISC-V vector ISA extension has been developed, comprising four vector-based instructions tailored to optimize the integration of the DIMC unit into the execution pipeline. These instructions enable:

\begin{itemize}
    \item High bandwidth loading of feature and weights data from the VRF to the DIMC;
    \item Seamless execution of MAC operations within the DIMC unit, with full control over configuration parameters;
    \item Optimized handling and storage of partial and final results with minimal memory overhead.
\end{itemize}

Isolating the DIMC as a functional unit enables parallelism with vector units, avoids access conflicts, reduces memory traffic, and removes coherence issues by routing all exchanges through the VRF.

\subsection{Proposed RISC-V Custom Instructions}

To integrate the DIMC accelerator efficiently into the RISC-V vector pipeline, we propose a small set of specialized vector instructions. These instructions are organized into two distinct functional groups based on their roles:

\begin{itemize}
\item Data Load Instructions (\texttt{DL.I}, \texttt{DL.M}) — manage high-bandwidth transfers respectively from the VRF to the Input Buffer and DIMC Memory.
\item Compute and Write-Back Instructions (\texttt{DC.P}, \texttt{DC.F}) — trigger MAC operations inside DIMC, handle the associated configuration signals, and manage the efficient write-back of partial or final results to the VRF.
\end{itemize}

\begin{figure}[!t]
    \centering
    \includegraphics[width=\columnwidth]{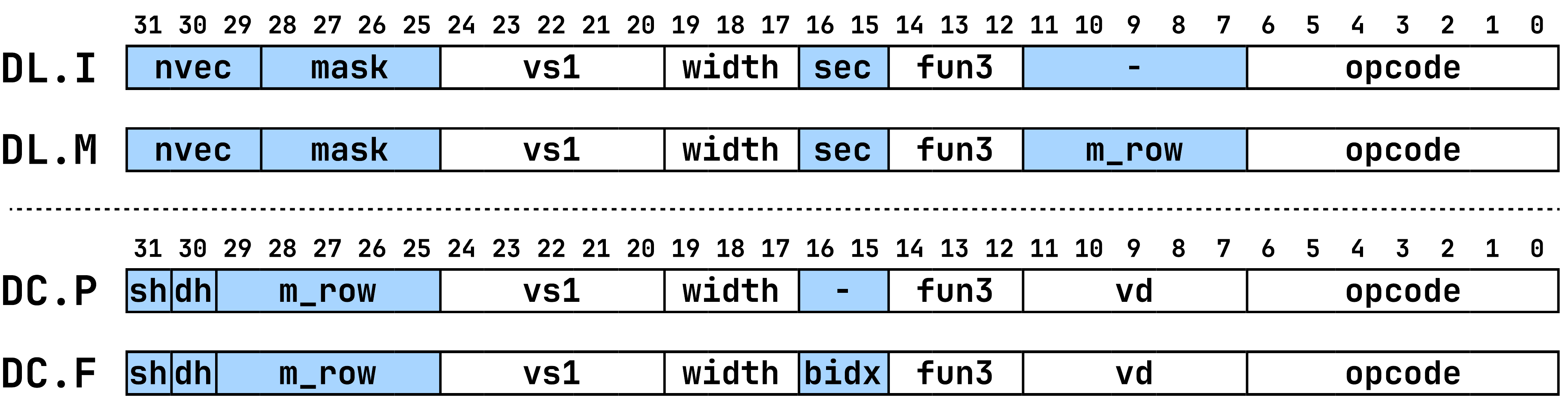}
    \caption{Bit-level encoding of the proposed custom instructions for DIMC accelerator integration. Customized bit fields are highlighted in blue.}
    \label{fig:custom_instructions}
\end{figure}

The bit-level encoding of these custom instructions is detailed in Fig.~\ref{fig:custom_instructions}, where customized fields are highlighted in blue.

\noindent\textbf{DIMC Input Buffer Load (DL.I).}
The \texttt{DL.I} instruction transfers between 64 and 256 bits of data from the Vector Register File (VRF) to the DIMC’s input buffer. It reads data from \texttt{nvec} consecutive VRF registers, starting at \texttt{vs1}, using a valid-bit \texttt{mask}. The loaded data is written into one of the four 256-bit sectors, specified by \texttt{sec}, within the DIMC’s 1024-bit input buffer.

\noindent\textbf{DIMC Memory Load (DL.M).}
The \texttt{DL.M} instruction operates similarly to \texttt{DL.I}, but includes an additional field, \texttt{m\_row}, which specifies the DIMC memory row where the data should be loaded.

\noindent\textbf{DIMC Compute \& Partial Sum Store (DC.P).}
The \texttt{DC.P} instruction performs an in-memory MAC operation between the data in the DIMC input buffer and the weights stored in the specified memory row (\texttt{m\_row}). It takes a 24-bit partial sum as input from the half of the \texttt{vs1} register selected by \texttt{sh}, and stores the resulting 24-bit partial result into the half of the \texttt{vd} register specified by \texttt{dh}.

\noindent\textbf{DIMC Compute \& Final Sum Store (DC.F).}
The \texttt{DC.F} instruction performs the same in-memory MAC operation as \texttt{DC.P}, but additionally applies the ReLU activation function and stores the final quantized result in compact form. The result is written into a specific byte of the half of the \texttt{vd} register in the VRF, selected by the \texttt{dh} (half selector) and \texttt{bidx} (byte index) fields.

\vspace{5pt}

Partial sums generated by \texttt{DPS} are padded from 24 bits to 32 bits for VRF alignment, with possible future optimizations to reduce overhead. Final outputs from \texttt{DSS}, quantized to 1-, 2-, or 4-bit precision, are padded to 4 bits and efficiently packed into bytes—two 4-bit results per byte. In the case of an odd number of results, the last half-byte remains unused, defining the maximum padding overhead.

These custom instructions collectively provide a concise yet powerful ISA extension, enabling streamlined integration and operation of the DIMC accelerator within a vector RISC-V core. From an encoding perspective, the four instructions are mapped to the RISC-V custom-0 space, which is explicitly reserved for non-standard extensions. This guarantees that no conflicts arise with current vector extensions, and sufficient encoding space remains available to accommodate further custom instructions in future work.

\section{Experimental Results}
This section outlines the experimental setup and the approach used to evaluate the performance of the DIMC-enhanced RVV processor against the baseline core. The results of the comparison highlight three key achievements: (1) the high utilization of the DIMC tile enabling near-peak computational throughput, (2) the significant speedup achieved with and without area normalization, and (3) the robust performance sustained even when DIMC hardware capacity limits are exceeded. We focus on comparison with the baseline RVV core because the objective of this work is to demonstrate the incremental benefit that DIMC integration brings to an established industrial-grade architecture. In other words, the emphasis is on quantifying the performance gain achievable by directly embedding DIMC into the vector pipeline, rather than on outperforming unrelated accelerator designs.

\subsection{Experimental Setup}

The evaluation was conducted using a cycle-approximate simulation framework developed 
specifically for the DIMC-enhanced RVV processor. This custom simulator models instruction-level execution with timing granularity sufficient to capture the interplay between the core pipeline and the tightly coupled DIMC unit.

The simulation tool generates execution traces from custom instruction streams derived from deep learning models. Each instruction is assigned a latency based on the hardware pipeline structure and stall conditions. The simulator explicitly models:
\begin{itemize}
    \item Instruction latencies, including vector loads, stores, and arithmetic operations;
    \item Pipeline stalls and execution flow control;
    \item Custom DIMC instruction timing, which reflects the internal datapath latency and tightly coupled access to the registers.
\end{itemize}

Area estimates for both the baseline and DIMC-enhanced architectures were obtained through RTL synthesis using Cadence tools, targeting a P18 CMOS 
process node from STMicroelectronics. These values are used to compute area-normalized performance metrics. While energy measurements are not reported in this work, future iterations may include RTL-based power estimates or model-based energy approximations.

The benchmark used for evaluation is ResNet50, from which individual convolutional and fully connected layers were extracted and translated into instruction streams. Each layer is mapped to the DIMC-enhanced RVV processor through a toolchain that:
\begin{enumerate}
\item Load kernel weights into the DIMC memory (up to 32 kernels).
\item Load one patch of feature data into the DIMC input buffer.
\item Trigger MAC operations using custom compute instructions.
\item Slide input window across the feature map and repeat 2–3.
\item Reload kernels if needed and continue the iteration.
\end{enumerate}

All MAC operations are executed inside the DIMC, with the RVV core orchestrating data movement and issuing compute instructions via the extended ISA. The simulator itself is not publicly released due to industrial confidentiality constraints.

\textbf{Assumptions.}
The following assumptions were made in the simulations to simplify the analysis, clearly define experimental boundaries, and avoid the complexities associated with advanced data mapping strategies and compiler modifications:

\begin{itemize}    
\item \textit{Vector Instruction Issuing}: Simulations did not consider double-issue vector instruction execution, simplifying modeling at the expense of capturing peak theoretical performance.

\item \textit{Memory Access Latency}: Although memory access is not modeled cycle-by-cycle, a fixed-latency external memory is assumed. This is sufficient for our analysis, since all data exchanges with the DIMC are tightly coupled and do not involve DMA.
    
\item \textit{Data Fetching and Reuse}: Simulations involving DIMC assumed that each feature map was loaded directly from memory, even in cases where data reuse could reduce memory accesses. This conservative assumption leads to sub-optimal results.

\item \textit{Resolution Limitation}: The baseline RVV architecture supports a minimum data resolution of 8 bits, while the DIMC unit used is limited to a maximum of 4 bits.
    
\item \textit{DIMC Capacity Constraint}: DIMC-supported convolutions were limited to kernels where the total bits per single channel did not exceed 1024 bits.
    
\item \textit{Layer Type Acceleration:} DIMC acceleration specifically targets convolutional and fully connected layers. Operations such as pooling, which rely on the standard RVV ISA, yield similar performance across both baseline and DIMC-enhanced architectures and were thus excluded from simulations results.  

\end{itemize}

\textbf{Performance Evaluation Metrics.} We evaluated the following three performance metrics:
\begin{enumerate}
    \item OPs (Operations Per Second): The total number of operations executed per second, calculated based on a 500 MHz clock frequency. This metric provides a direct measure of the system’s computational throughput.
    \item Speedup: The ratio of clock cycles executed by the baseline RVV processor compared to the DIMC-enhanced version, quantifying the relative performance improvement:
    \[Speedup = \frac{\text{ClockCycles}_{\text{Baseline RVV}}}{\text{ClockCycles}_{\text{DIMC RVV}}}\]
    \item Area-Normalized Speedup: The speedup metric normalized with respect to the baseline RVV area, allowing performance improvements to be evaluated in the context of increased hardware complexity.
    \[ANS = {Speedup}*{\frac{Area_{Baseline}}{Area_{DIMC RVV}}}\]
\end{enumerate}

\begin{figure}[t]
    \centering
    \includegraphics[width=\columnwidth]{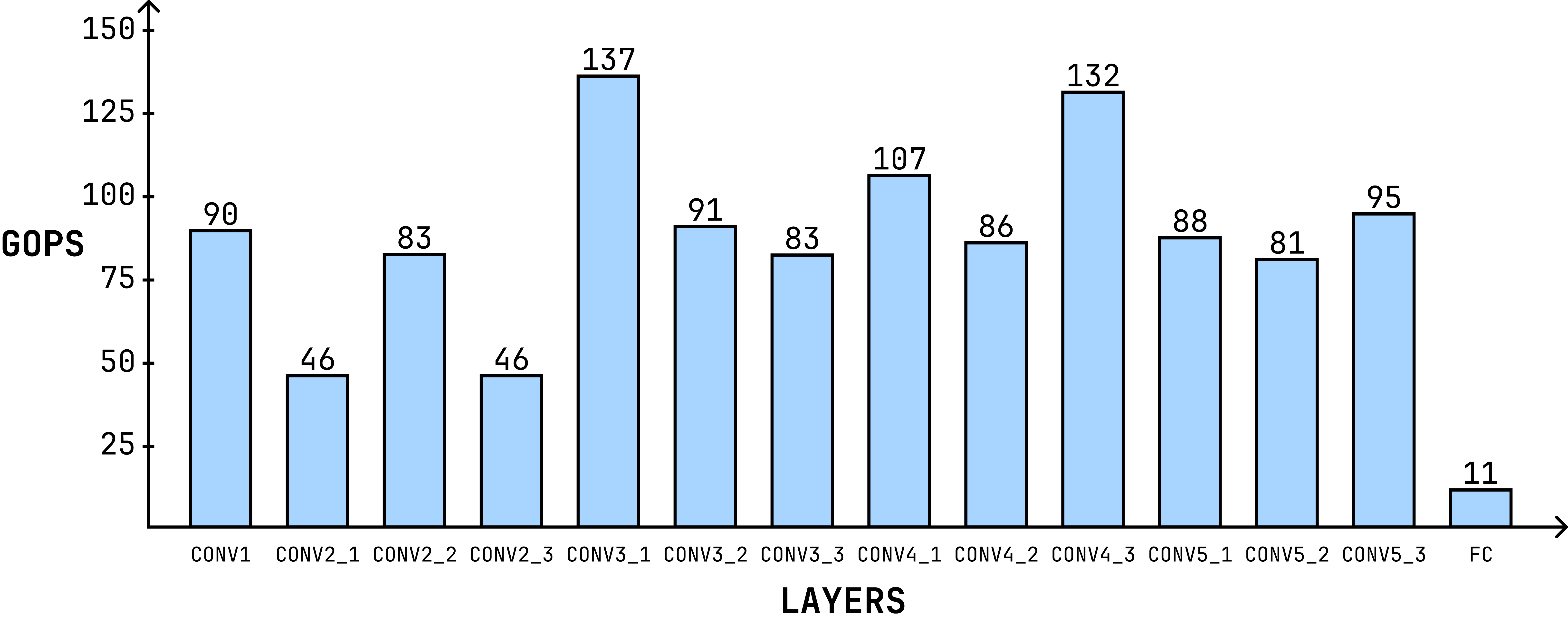}
    \caption{GOPS Achieved per Layer in ResNet50}
    \label{fig:gops}
\vspace{6pt}
    \centering
    \includegraphics[width=\columnwidth]{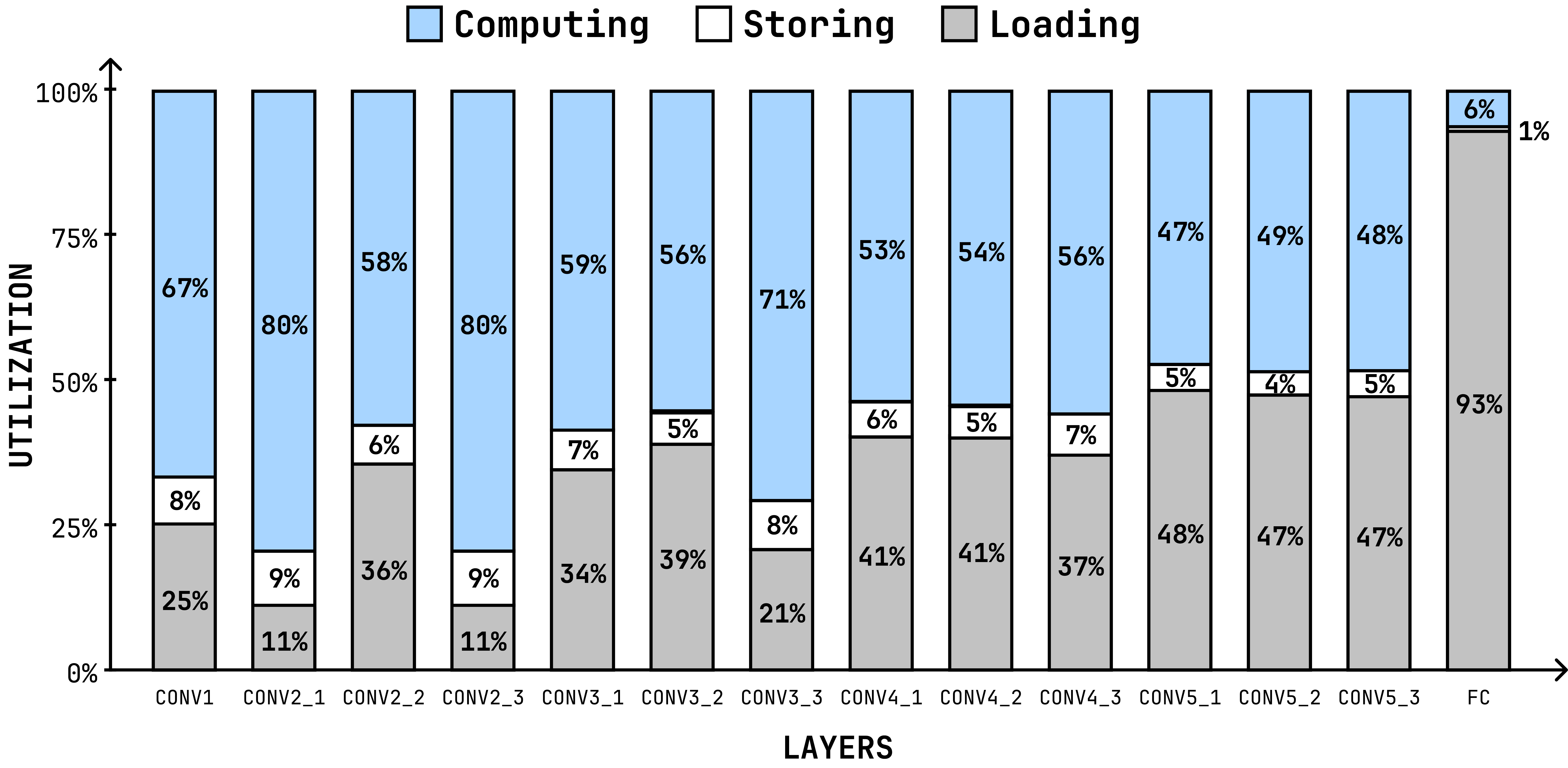}
    \caption{Operation Distribution (Computing, Loading, Storing) per Layer in ResNet50}
    \label{fig:distribution}
\vspace{6pt}
    \centering
    \includegraphics[width=\columnwidth]{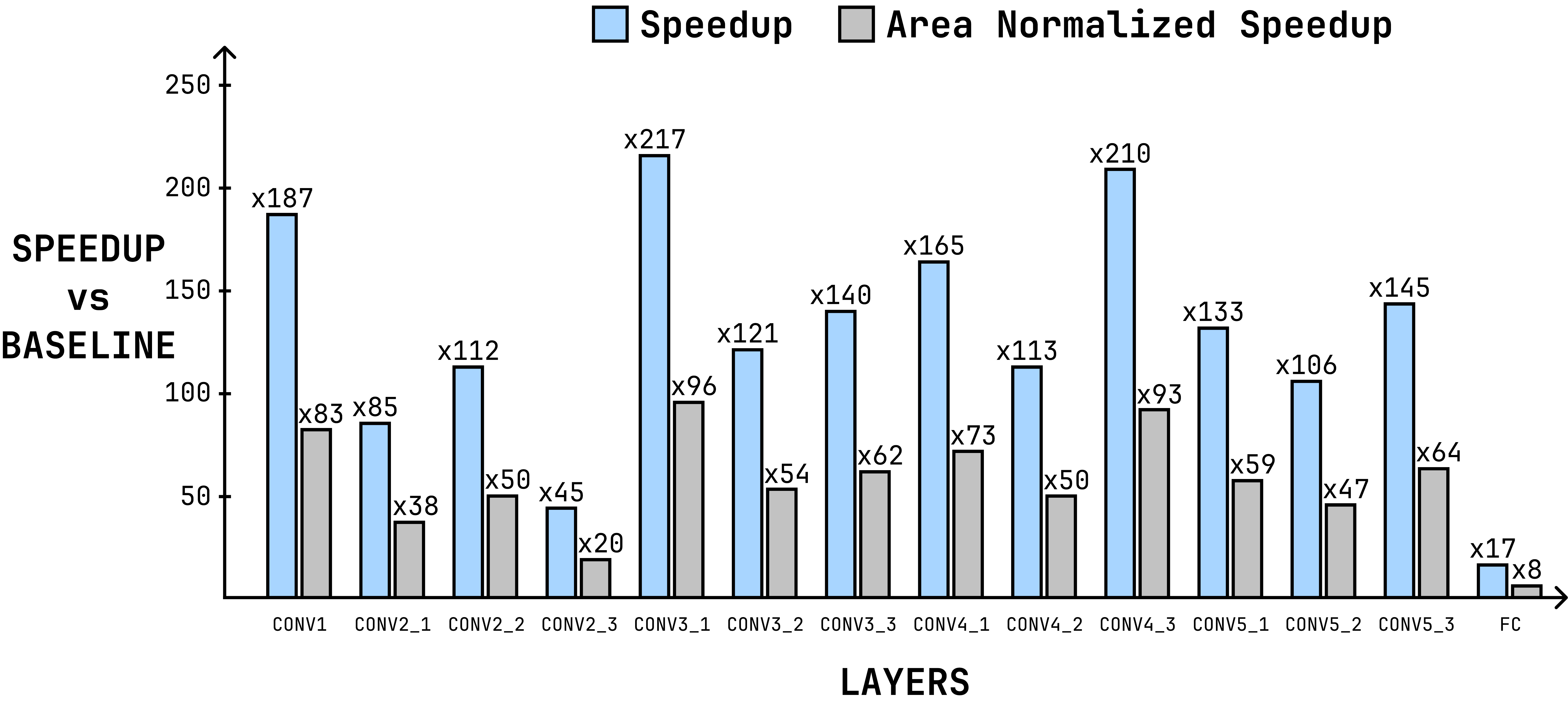}
    \caption{Speedup and Area-Normalized Speedup per Layer in ResNet50}
    \label{fig:speedup}
\end{figure}

\begin{table*}[htbp]
\caption{Comparison of IMC-Integrated RISC-V Architectures}
\centering
\resizebox{\textwidth}{!}{%
\label{tab:imc_compare}
\scriptsize
\renewcommand{\arraystretch}{1.1}
\begin{tabular}{|l|c|c|c|c|c|c|c|c|}
\hline
 & \textbf{Scalar/Vector} & \textbf{Integration} & \textbf{Memory type} & \textbf{Memory size} & \textbf{Freq. [MHz]}       & \textbf{Reported Perf.} & \textbf{Norm. GOPS\footnotemark[1]} \\ \hline
CIMR-V \cite{guo_cimr-v_2024}  & Scalar & Loose  & 10T SRAM \cite{9634797} & 64 KB  & 50                                    & 26.2 TOPS @INT1 & $\sim$2.6 TOPS @INT4, 500 MHz* \\ \hline
AI-PiM \cite{verma_ai-pimextending_2022} & Scalar & Tight (In-Pip.) & 8T SRAM \cite{9062985} & 500 B & --                   & -- & -- \\ \hline
VPU-CIM \cite{j_vpu-cim_2024} & Vector & Loose  & RRAM & 8 KB & 25                                                          & -- & -- \\ \hline
Vecim \cite{wang_306_2024} & Vector & Tight  & 8T SRAM & -- & 250                                                           & 31.8 GOPS @INT8 & $\sim$63.6 GOPS @INT4, 500 MHz* \\ \hline
RDCIM \cite{10403109} & Scalar & Tight  & 8T SRAM & 64 KB & 200                                                             & -- & -- \\ \hline
\textbf{This Work} & \textbf{Vector} & \textbf{Tight (In-Pip.)} & \textbf{8T SRAM} & \textbf{4 KB} & \textbf{500}           & \textbf{137 GOPS @INT4} & \textbf{137 GOPS @INT4} \\ \hline
\end{tabular}
}
\end{table*}


\subsection{Efficient Utilization of the DIMC Tile}

To evaluate the computational throughput achieved by the DIMC accelerator, simulations were performed on each convolutional and fully connected layer in ResNet50 \cite{7780459}. As shown in Fig.~\ref{fig:gops}, the DIMC-enhanced processor reaches peak throughput values close to its theoretical limit (based on MAC unit performance and assuming no loading or storing penalties), achieving over 100 GOPS in many layers and peaking at 137 GOPS.

This high performance is explained by the efficient scheduling and utilization of the DIMC tile, thanks to the new custom instructions obtained through the extension of the instruction set architecture. The breakdown of operations per layer in Fig.~\ref{fig:distribution} confirms that the DIMC spends the majority of execution time on computation rather than data loading or result storing, validating its ability to exploit hardware resources effectively. This result highlights the value of the In-Pipeline integration design in a Vector RISC-V core, which enables a high compute operation ratio.

\subsection{Speedup over Baseline RVV Core}

The integration of DIMC leads to a substantial speedup over the baseline RVV core. As shown in Fig.~\ref{fig:speedup}, across all ResNet50 layers, the DIMC-enhanced architecture consistently outperforms the baseline, with raw speedup values exceeding 200× in some layers and area-normalized speedup maintaining values well above 50×. These results demonstrate that even when accounting for the additional hardware cost, the tightly integrated DIMC provides significant performance advantages for edge AI workloads.

\footnotetext[1]{Normalized GOPS values with our work's precision and frequency.}

\subsection{Analysis of the Accelerated Workload}

To assess flexibility and generalization, we analyzed over 450 convolutional layers from models including  AlexNet \cite{krizhevsky_imagenet_2012}, VGG16 \cite{simonyan_very_2015}, ResNet \cite{7780459}, Inception \cite{szegedy_going_2014}, DenseNet \cite{huang_densely_2018}, EfficientNet \cite{tan_efficientnet_2020}, and MobileNet \cite{howard_mobilenets_2017}. Covering a broad range of feature maps and kernel dimensions, these configurations represent real-world scenarios. Across them, the DIMC-augmented system consistently outperforms the baseline, demonstrating high versatility.

While Table~\ref{tab:imc_compare} compares representative architectures, our evaluation focuses on the baseline RVV core to isolate the impact of DIMC integration.
Unlike prior works limited to layers fitting architectural constraints, our analysis includes configurations that exceed system limits. This stresses the flexibility of the system in scenarios requiring frequent kernel switching and sub-optimal memory mapping.
The two primary architectural constraints are:
\begin{itemize}
    \item a maximum single-kernel bitwidth of 1024 bits (requires \textit{tiling});
    \item a limit of 32 kernel in DIMC memory (requires \textit{grouping}).
\end{itemize}

\begin{figure}[t]
    \centering
    \includegraphics[width=\columnwidth]{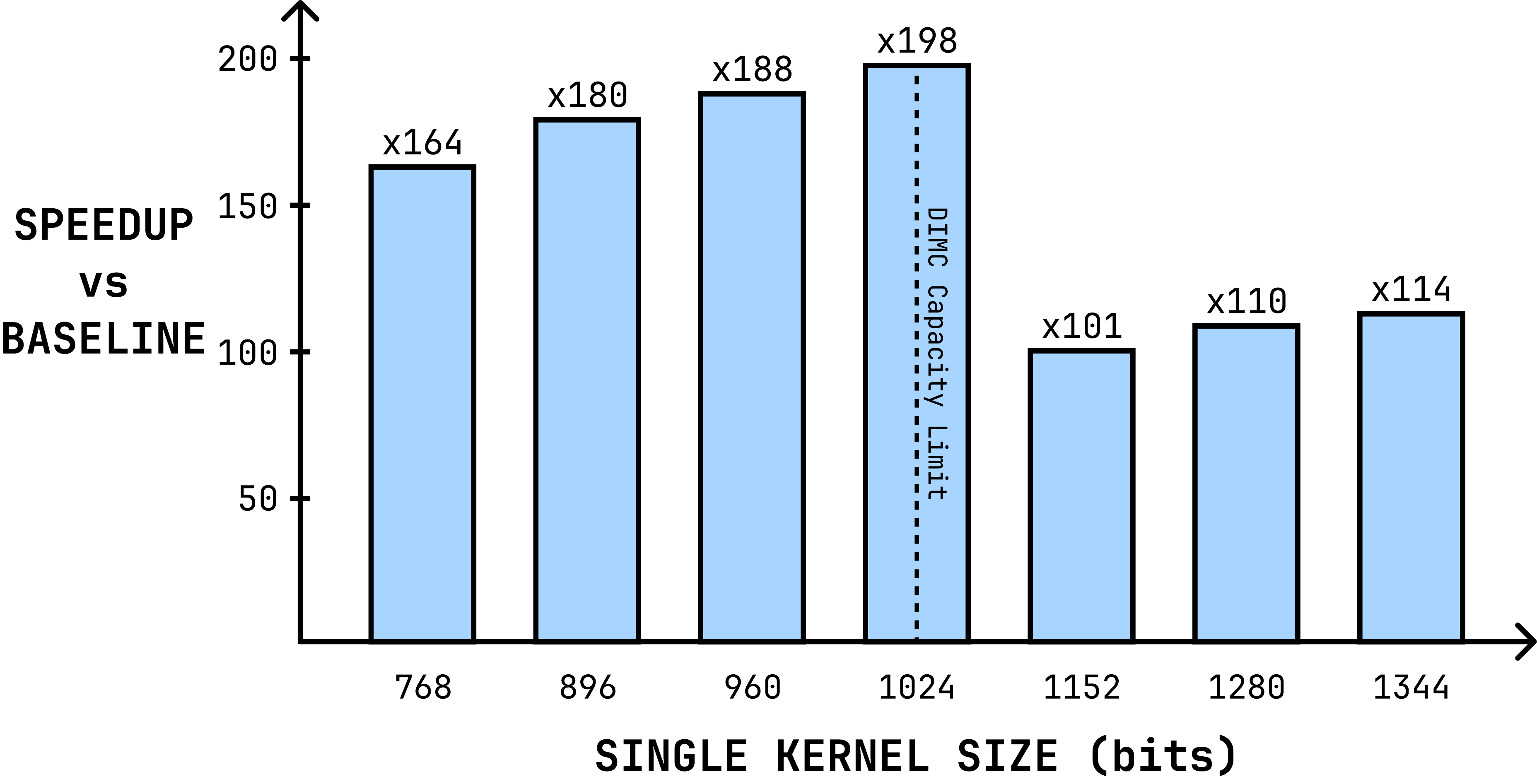}
    \caption{Speedup Degradation due to tiling (OCH=32, KH=2, KW=2)}
    \label{fig:batch}
\vspace{20pt}
    \includegraphics[width=\columnwidth]{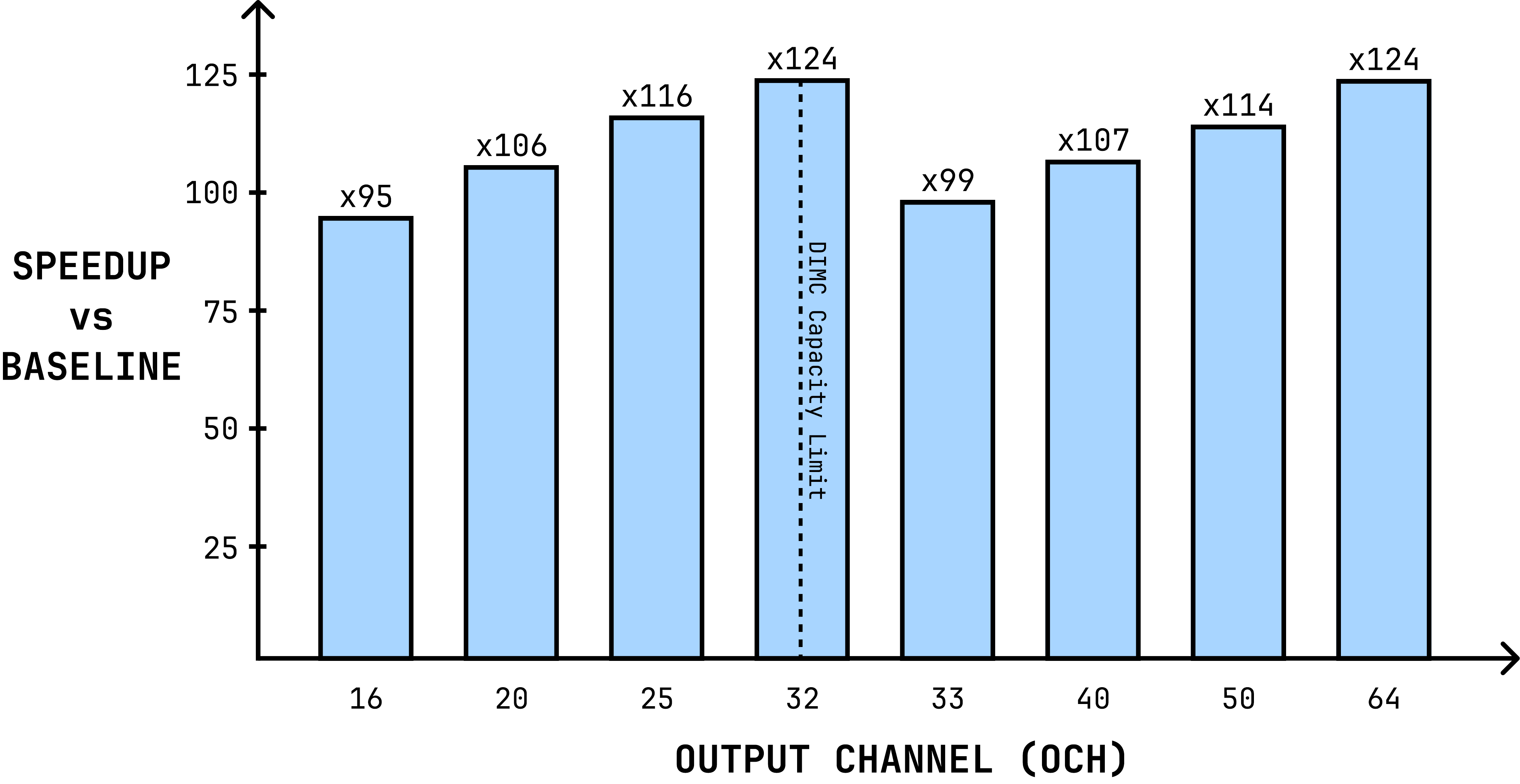}
    \caption{Speedup Degradation due to grouping (ICH=32, KH=2, KW=2)}
    \label{fig:kbatch}
\end{figure}

Fig.\ref{fig:batch} shows how \textit{tiling} is employed when the kernel size exceeds the 1024-bit threshold. While this incurs a performance drop due to serial loading and computation, the DIMC architecture still maintains a strong advantage over the baseline. 
Similarly, Fig.\ref{fig:kbatch} evaluates the case of \textit{grouping}. Despite the forced segmentation of compute, the architecture sustains notable speedup.

These results underline not just the high baseline throughput of the DIMC tiles, but more importantly, the architectural strategy introduced in this work. The combination of a streamlined custom instruction set (DL.I, DL.M, DC.P, and DC.F) and its tight integration with the RISC-V Vector Extension enables precise control over kernel loading, compute scheduling, and data reuse. This coordination is what allows the system to adapt dynamically to sub-optimal scenarios—such as fragmented kernel shapes or frequent weight switching—without losing efficiency. Rather than tailoring workloads to fit static hardware constraints, the proposed architecture adapts to the workload, sustaining high tile utilization and acceleration across diverse conditions. This makes it a practical and robust solution for applications with variable and unpredictable inference patterns.

\subsection{Summary and Comparison with Prior Architectures}

Finally, to contextualize the achieved results, Table~\ref{tab:imc_compare} compares the proposed architecture against representative IMC-integrated RISC-V processors, highlighting differences in integration strategy, core type, and peak performance.

Unlike prior works, this is the first design to tightly integrate a DIMC unit within a vector core pipeline as an FU. It achieves the highest reported performance (137 GOPS at INT4) while operating at 500 MHz with only 4 KB of DIMC memory. Compared to scalar or loosely coupled designs, our architecture offers superior throughput and control with minimal memory overhead.

\section{Conclusions}

This work demonstrates the first in-pipeline integration of a DIMC unit within a RISC-V vector processor, supported by a custom ISA extension for fine-grained control of dataflow and computation. By extending an industrial-grade vector core with a single DIMC tile, we show that the proposed integration model reaches up to 137 GOPS and more than 200× speedup over the baseline, with over 50× gain even when area-normalized. Focusing on a single tile allowed us to precisely define how a vector core can efficiently manage and exploit DIMC units, laying the foundation for future scaling to multiple tiles. Unlike prior designs that require workload tailoring, our architecture maintains high performance across diverse computations, making it a scalable and adaptive accelerator for next-generation edge AI.

\bibliographystyle{IEEEtran}
\bibliography{biblio.bib}

\end{document}